\documentclass[aip,reprint,author-year,nofootinbib,floatfix]{revtex4-2}

\usepackage{physjour_aas_macros}

\usepackage{graphicx}
\usepackage{amsmath,amssymb}             
\usepackage{stmaryrd}					
\usepackage[utf8]{inputenc}
\usepackage[T1]{fontenc}
\usepackage[dvipsnames]{xcolor}
\usepackage{color}
\usepackage[raggedright]{titlesec}
\usepackage[normalem]{ulem}
\definecolor{darkblue}{rgb}{0.0,0.0,0.4}
\definecolor{darkred}{rgb}{0.7,0.0,0.0}
\definecolor{darkgreen}{rgb}{0.0,0.5,0.0}
\definecolor{C0}{HTML}{1f77b4}
\definecolor{C1}{HTML}{ff7f0e}
\definecolor{C2}{HTML}{2ca02c}
\definecolor{C3}{HTML}{d62728}
\definecolor{C4}{HTML}{9467bd}
\definecolor{C5}{HTML}{8c564b}
\definecolor{C6}{HTML}{e377c2}
\definecolor{C7}{HTML}{7f7f7f}
\definecolor{C8}{HTML}{bcbd22}
\definecolor{C9}{HTML}{17becf}

\usepackage{tikz}
\usetikzlibrary{shapes,arrows,shadows}
\usepackage[export]{adjustbox}

\usepackage{scalerel}
\usetikzlibrary{svg.path}
\definecolor{orcidlogocol}{HTML}{A6CE39}
\tikzset{
  orcidlogo/.pic={
    \fill[orcidlogocol] svg{M256,128c0,70.7-57.3,128-128,128C57.3,256,0,198.7,0,128C0,57.3,57.3,0,128,0C198.7,0,256,57.3,256,128z};
    \fill[white] svg{M86.3,186.2H70.9V79.1h15.4v48.4V186.2z}
                 svg{M108.9,79.1h41.6c39.6,0,57,28.3,57,53.6c0,27.5-21.5,53.6-56.8,53.6h-41.8V79.1z M124.3,172.4h24.5c34.9,0,42.9-26.5,42.9-39.7c0-21.5-13.7-39.7-43.7-39.7h-23.7V172.4z}
                 svg{M88.7,56.8c0,5.5-4.5,10.1-10.1,10.1c-5.6,0-10.1-4.6-10.1-10.1c0-5.6,4.5-10.1,10.1-10.1C84.2,46.7,88.7,51.3,88.7,56.8z};
  }
}
\newcommand\orcid[1]{\href{https://orcid.org/#1}{\mbox{\scalerel*{
\begin{tikzpicture}[yscale=-1,transform shape]
\pic{orcidlogo};
\end{tikzpicture}
}{|}}}}

\usepackage{hyperref}
\hypersetup{colorlinks,breaklinks,
            linkcolor=darkblue,urlcolor=darkblue,
            anchorcolor=darkblue,citecolor=RoyalBlue}

\usepackage[mathscr]{euscript}
\usepackage{upgreek}
\DeclareFontFamily{OT1}{pzc}{}
\DeclareFontShape{OT1}{pzc}{m}{it}{<-> s * [1.10] pzcmi7t}{}
\DeclareMathAlphabet{\mathpzc}{OT1}{pzc}{m}{it}

\titleformat{\section}{\selectfont \normalfont\raggedright\sffamily\small\bfseries\uppercase}{\thesection.}{1em}{}{}
\titleformat{\subsection}{\selectfont \normalfont\raggedright\sffamily\small\bfseries}{\thesubsection.}{1em}{}{}
\titleformat{\subsubsection}{\selectfont \normalfont\sffamily\small\bfseries}{\thesubsection.\thesubsubsection}{1em}{}{}
\titleformat{\paragraph}[runin]{\selectfont \normalfont\sffamily\small\bfseries}{\thesubsection.\thesubsubsection.\theparagraph}{1em}{}[.]

\usepackage{tikz}
\usetikzlibrary{shapes,arrows,shadows}
\usepackage[export]{adjustbox}

\newcommand{\deriv}[2]{\dfrac{\mathrm{d} #1}{\mathrm{d} #2}}

\newcommand{\p}{\mathpzc{P}}

\newcommand{\selfi}{\textsc{selfi}}

\begin{document}

\title{Simulation-based inference of Bayesian hierarchical models while checking for model misspecification}



\author{Florent~Leclercq}
\email{florent.leclercq@iap.fr}
\homepage{https://www.florent-leclercq.eu/}
\thanks{ORCID: \orcid{0000-0002-9339-1404} \href{https://orcid.org/0000-0002-9339-1404}{0000-0002-9339-1404}}
\affiliation{CNRS \& Sorbonne Université, UMR 7095, Institut d’Astrophysique de Paris, 98 bis boulevard Arago, F-75014 Paris, France}


\date{\today}

\begin{abstract}
\noindent This paper presents recent methodological advances to perform simulation-based inference (SBI) of a general class of Bayesian hierarchical models (BHMs), while checking for model misspecification. Our approach is based on a two-step framework. First, the latent function that appears as second layer of the BHM is inferred and used to diagnose possible model misspecification. Second, target parameters of the trusted model are inferred via SBI. Simulations used in the first step are recycled for score compression, which is necessary to the second step. As a proof of concept, we apply our framework to a prey-predator model built upon the Lotka-Volterra equations and involving complex observational processes.
\end{abstract}

\maketitle



\section{Introduction}
\label{sec:Introduction}

Model misspecification is a long-standing problem for Bayesian inference: when the model differs from the actual data-generating process, posteriors tend to be biased and/or overly concentrated. 
In this paper, we are interested the problem of model misspecification for a particular, but common, class of Bayesian hierarchical models (BHMs): those that involve a latent function, such as the primordial power spectrum in cosmology \citep[e.g.][]{Leclercq2019SELFI} or the population model in genetics \citep[e.g.][]{Rousset2007}. 

Simulation-based inference (SBI) only provides the posterior of top-level target parameters and marginalises over all other latent variables of the BHM. 
Alone, it is therefore unable to diagnose whether the model is misspecified. 
Key insights regarding the issue of model misspecification can usually be obtained from the posterior distribution of the latent function, as there often exists an independent theoretical understanding of its values. 
An approximate posterior for the latent function (a much higher-dimensional quantity than the target vector of parameters) can be obtained using {\selfi} \citep[Simulator expansion for likelihood-free inference,][]{Leclercq2019SELFI}, an approach based on the likelihood of an alternative parametric model, constructed by linearising model predictions around an expansion point.

This paper presents a framework that combines {\selfi} and SBI while recycling the necessary simulations. 
The simulator is first linearised to obtain the {\selfi} posterior of the latent function. 
Then, the same simulations are used for data compression to the score function (the gradient of the log-likelihood with respect to the parameters) and the final SBI posterior of target parameters is obtained. 

\section{Method}
\label{sec:Method}

\subsection{Bayesian hierarchical models with a latent function}

In this paper, we assume given a BHM consisting of the following variables: $\boldsymbol{\upomega} \in \mathbb{R}^N$ (vector of $N$ target parameters), $\boldsymbol{\uptheta} \in \mathbb{R}^S$ (vector containing the values of the latent function $\theta$ at $S$ support points), $\boldsymbol{\Phi} \in \mathbb{R}^P$ (data vector of $P$ components), and $\widetilde{\boldsymbol{\upomega}} \in \mathbb{R}^N$ (compressed data vector of size $N$). We typically expect $N \sim \mathcal{O}(5-10)$ target parameters, $S \sim \mathcal{O}(10^2-10^3)$ support points; $P$ can be any number and as large as $\mathcal{O}(10^7)$ for complex data models. 
We further assume that $\boldsymbol{\upomega}$ and $\boldsymbol{\uptheta}$ are linked by a deterministic function $\mathpzc{T}$, usually theoretically well-understood and numerically cheap. 
Therefore, the expensive and potentially misspecified part of the BHM is the probabilistic simulator linking the latent function $\boldsymbol{\uptheta}$ to the data $\boldsymbol{\Phi}$, $\p(\boldsymbol{\Phi}|\boldsymbol{\uptheta})$. The deterministic compression step $\mathpzc{C}$ linking $\boldsymbol{\Phi}$ to $\boldsymbol{\widetilde{\upomega}}$ is discussed later in section \ref{ssec:Score compression and simulation-based inference}. 

\subsection{Latent function inference with SELFI}

The first part of the framework proposed in this paper is to infer the latent function $\boldsymbol{\uptheta}$ conditional on observed data $\boldsymbol{\Phi}_\mathrm{O}$. This is an inference problem in high dimension ($S$, the number of support points for the latent function $\theta$) which means that usual SBI frameworks, allowing a general exploration of parameter space, will fail and that stronger assumptions are required. {\selfi} \citep{Leclercq2019SELFI} relies upon the simplification of the inference problem around an expansion point $\boldsymbol{\uptheta}_0$.

The first assumption is a Taylor expansion (linearisation) of the mean data model around $\boldsymbol{\uptheta}_0$. Namely, if $\boldsymbol{\hat{\Phi}}_{\boldsymbol{\uptheta}} \equiv \mathrm{E} \left[ \boldsymbol{\Phi}_{\boldsymbol{\uptheta}} \right]$ is the expectation value of $\boldsymbol{\Phi}_{\boldsymbol{\uptheta}}$, where $\boldsymbol{\Phi}_{\boldsymbol{\uptheta}}$ are simulations of $\boldsymbol{\Phi}$ given $\boldsymbol{\uptheta}$ (i.e. $\boldsymbol{\Phi}_{\boldsymbol{\uptheta}} \curvearrowleft \p(\boldsymbol{\Phi}|\boldsymbol{\uptheta})$), we assume that
\begin{equation}
\boldsymbol{\hat{\Phi}}_{\boldsymbol{\uptheta}} \approx \textbf{f}_0 + \nabla \textbf{f}_0 \cdot (\boldsymbol{\uptheta}-\boldsymbol{\uptheta}_0) \equiv \textbf{f}(\boldsymbol{\uptheta}),
\label{eq:linearised_black_box}
\end{equation}
where $\textbf{f}_0 \equiv \boldsymbol{\hat{\Phi}}_{\boldsymbol{\uptheta}_0}$ is the mean data model at the expansion point $\boldsymbol{\uptheta}_0$, and $\nabla \textbf{f}_0$ is the gradient of $\textbf{f}_0$ at the expansion point (for simplification we note $\nabla \textbf{f}_0 = \nabla_{\boldsymbol{\uptheta}} \textbf{f}_0$, where the gradient is taken with respect to $\boldsymbol{\uptheta}$). The second assumption is that the (true) implicit likelihood of the problem is replaced by a Gaussian effective likelihood: $\p(\boldsymbol{\Phi}_\mathrm{O}|\boldsymbol{\uptheta}) \equiv \exp\left[ \hat{\ell}_{\boldsymbol{\uptheta}}(\boldsymbol{\uptheta}) \right]$ with
\begin{equation}
-2 \hat{\ell}_{\boldsymbol{\uptheta}}(\boldsymbol{\uptheta}) \approx \log\left| 2\pi\textbf{C}_0 \right| + \left[\boldsymbol{\Phi}_\mathrm{O} - \textbf{f}(\boldsymbol{\uptheta})\right]^\intercal \textbf{C}_0^{-1} \left[\boldsymbol{\Phi}_\mathrm{O} - \textbf{f}(\boldsymbol{\uptheta})\right],
\label{eq:linearised_effective_likelihood}
\end{equation}
where $\textbf{C}_0$ is the data covariance matrix at the expansion point $\boldsymbol{\uptheta}_0$.

The {\selfi} framework is fully characterised by $\textbf{f}_0$, $\textbf{C}_0$, and $\nabla \textbf{f}_0$, which, if unknown, can be evaluated through forward simulations only. The numerical computation requires $N_0$ simulations at the expansion point (to evaluate the empirical mean $\textbf{f}_0$ and empirical covariance matrix $\textbf{C}_0$), and $N_s$ simulations in each direction of parameter space (to evaluate the empirical gradient $\nabla \textbf{f}_0$ via first-order forward finite differences). The total is $N_0 + N_s \times S$ simulations. $N_0$ and $N_s$ should be of the order of the dimensionality of the data space $P$, giving a total cost of $\mathcal{O}(\gtrsim P(S+1))$ model evaluations.

To fully characterise the Bayesian problem, one requires a prior on $\boldsymbol{\uptheta}$, $\p(\boldsymbol{\uptheta})$. Any prior can be used if one is ready to use numerical techniques to explore the posterior (such as standard Markov Chain Monte Carlo), using the linearised data model and Gaussian effective likelihood. However, a remarkable analytic result with {\selfi} is that, if the prior is Gaussian with a mean equal to the expansion point $\boldsymbol{\uptheta}_0$, i.e.
\begin{equation}
-2\log \p(\boldsymbol{\uptheta}) \equiv \log\left| 2\pi \textbf{S} \right| + (\boldsymbol{\uptheta}-\boldsymbol{\uptheta}_0)^\intercal \textbf{S}^{-1} (\boldsymbol{\uptheta} - \boldsymbol{\uptheta}_0) ,
\label{eq:prior}
\end{equation}
then the effective posterior is also Gaussian:
\begin{equation}
-2\log \p(\boldsymbol{\uptheta}|\boldsymbol{\Phi}_\mathrm{O}) \approx \log \left| 2\pi\boldsymbol{\Gamma} \right| + (\boldsymbol{\uptheta}-\boldsymbol{\upgamma})^\intercal \boldsymbol{\Gamma}^{-1} (\boldsymbol{\uptheta}-\boldsymbol{\upgamma}).
\end{equation}
The posterior mean and covariance matrix are given by
\begin{eqnarray}
\boldsymbol{\upgamma} & \equiv & \boldsymbol{\uptheta}_0 + \boldsymbol{\Gamma} \, (\nabla \textbf{f}_0)^\intercal \, \textbf{C}_0^{-1} (\boldsymbol{\Phi}_\mathrm{O}-\textbf{f}_0), \label{eq:filter_mean}\\
\boldsymbol{\Gamma} & \equiv & \left[ (\nabla \textbf{f}_0)^\intercal \, \textbf{C}_0^{-1} \nabla \textbf{f}_0 + \textbf{S}^{-1} \right]^{-1} \label{eq:filter_var}
\end{eqnarray}
\cite[see][appendix B, for a derivation]{Leclercq2019SELFI}. They are fully characterised by the expansion variables $\boldsymbol{\uptheta}_0$, $\textbf{f}_0$, $\textbf{C}_0$, and $\nabla \textbf{f}_0$, as well as the prior covariance matrix $\textbf{S}$.

\subsection{Check for model misspecification}

The {\selfi} posterior can be used as a check for model misspecification. Visually checking the reconstructed $\boldsymbol{\upgamma}$ and $\boldsymbol{\Gamma}$ can yield interesting insights, especially if the latent function has some properties (such as an expected shape, periodicity, etc.) to which the data model may be sensitive if misspecified (see section \ref{ssec:Check for model misspecification}).

If a quantitative check for model misspecification is desired, we propose to use the Mahalanobis distance between the reconstruction $\boldsymbol{\upgamma}$ and the prior distribution $\p(\boldsymbol{\uptheta})$, defined formally by
\begin{equation}
d_\mathrm{M}(\boldsymbol{\uptheta}, \boldsymbol{\uptheta}_0 | \textbf{S}) \equiv \sqrt{\left(\boldsymbol{\uptheta}-\boldsymbol{\uptheta}_0\right)^\intercal \textbf{S}^{-1} (\boldsymbol{\uptheta}-\boldsymbol{\uptheta}_0)}.
\label{eq:Mahalanobis_distance_mscheck}
\end{equation}
The value of $d_\mathrm{M}(\boldsymbol{\upgamma}, \boldsymbol{\uptheta}_0 | \textbf{S})$ for the {\selfi} posterior mean $\boldsymbol{\upgamma}$ can be compared to an ensemble of values of $d_\mathrm{M}(\boldsymbol{\uptheta}_{\boldsymbol{\upomega}}, \boldsymbol{\uptheta}_0 | \textbf{S})$ for simulated latent functions $\boldsymbol{\uptheta}_{\boldsymbol{\upomega}} = \mathpzc{T}(\boldsymbol{\upomega})$, where samples $\boldsymbol{\upomega}$ are drawn from the prior $\p(\boldsymbol{\upomega})$.

\subsection{Score compression and simulation-based inference}
\label{ssec:Score compression and simulation-based inference}

Having checked the BHM for model misspecification, we now address the second part of the framework, aiming at inferring top-level parameters $\boldsymbol{\upomega}$ given observations. 
SBI is known to be difficult when the dimensionality of the data space $P$ is high. For this reason, data compression is usually necessary. Data compression can be thought of as an additional layer at the bottom of the BHM, made of a deterministic function $\mathpzc{C}$ acting on $\boldsymbol{\Phi}$. In practical scenarios, data compression shall preserve as much information about $\boldsymbol{\upomega}$ as possible, meaning that compressed summaries $\mathpzc{C}(\boldsymbol{\Phi})$ shall be as close as possible to sufficient summary statistics of $\boldsymbol{\Phi}$, i.e. $\p(\boldsymbol{\upomega}|\mathpzc{C}(\boldsymbol{\Phi})) = \p(\boldsymbol{\upomega}|\boldsymbol{\Phi})$. 

Here, we propose to use score compression \citep{AlsingWandelt2018}. We make the assumption (for compression only, not for later inference) that $\p(\boldsymbol{\Phi}|\boldsymbol{\upomega})$ is Gaussian-distributed: $\p(\boldsymbol{\Phi}_\mathrm{O}|\boldsymbol{\upomega}) \equiv \exp \left[ \hat{\ell}_{\boldsymbol{\upomega}}(\boldsymbol{\upomega}) \right]$ where $\hat{\ell}_{\boldsymbol{\upomega}}(\boldsymbol{\upomega}) = \hat{\ell}_{\boldsymbol{\uptheta}}(\mathpzc{T}(\boldsymbol{\upomega}))$ (see equation \eqref{eq:linearised_effective_likelihood}).
The score function $\nabla_{\boldsymbol{\upomega}} \hat{\ell}_{\boldsymbol{\upomega}0}$ is the gradient of this log-likelihood with respect to the parameters $\boldsymbol{\upomega}$ at a fiducial point $\boldsymbol{\upomega}_0$ in parameter space. 
Using as fiducial point the values that generate the {\selfi} expansion point (i.e. $\boldsymbol{\upomega}_0$ such that $\boldsymbol{\uptheta}_0 = \mathpzc{T}(\boldsymbol{\upomega}_0)$), a quasi maximum-likelihood estimator for the parameters is $\widetilde{\boldsymbol{\upomega}}_\mathrm{O} \equiv \boldsymbol{\upomega}_0 + \textbf{F}^{-1}_0 \nabla_{\boldsymbol{\upomega}} \hat{\ell}_{\boldsymbol{\upomega}0}$, where the Fisher matrix $\textbf{F}_0$ and the gradient of the log-likelihood are evaluated at $\boldsymbol{\upomega}_0$. 
Compression of $\boldsymbol{\Phi}_\mathrm{O}$ to $\widetilde{\boldsymbol{\upomega}}_\mathrm{O}$ yields $N$ compressed statistics that are optimal in the sense that they preserve the Fisher information content of the data \citep{AlsingWandelt2018}. 

In our case, the covariance matrix $\textbf{C}_0$ is assumed not to depend on parameters ($\nabla_{\boldsymbol{\upomega}} \textbf{C}_0 = 0$), and the expression for $\mathpzc{C}(\boldsymbol{\Phi})$ is therefore
\begin{equation}
\mathpzc{C}(\boldsymbol{\Phi}) = \boldsymbol{\widetilde{\upomega}} \equiv \boldsymbol{\upomega}_0 + \textbf{F}^{-1}_0 \left[ (\nabla_{\boldsymbol{\upomega}} \textbf{f}_0)^\intercal \textbf{C}_0^{-1} (\boldsymbol{\Phi} - \textbf{f}_0) \right].
\label{eq:compression_mle}
\end{equation}
The Fisher matrix of the problem further takes a simple form:
\begin{equation}
\textbf{F}_0 \equiv -\mathrm{E}\left[ \nabla_{\boldsymbol{\upomega}} \nabla_{\boldsymbol{\upomega}} \hat{\ell}_{\boldsymbol{\upomega}0}(\boldsymbol{\upomega}) \right] = (\nabla_{\boldsymbol{\upomega}} \textbf{f}_0)^\intercal \textbf{C}_0^{-1} \nabla_{\boldsymbol{\upomega}} \textbf{f}_0.
\label{eq:Fisher_matrix}
\end{equation}
We therefore need to evaluate
\begin{equation}
\nabla_{\boldsymbol{\upomega}} \textbf{f}_0 = \nabla \textbf{f}_0 \cdot \left. \frac{\partial \mathpzc{T}(\boldsymbol{\upomega})}{\partial \boldsymbol{\upomega}} \right|_{\boldsymbol{\upomega} = \boldsymbol{\omega}_0}.
\label{eq:gradient_omega}
\end{equation}
Importantly, in equations \eqref{eq:compression_mle}, \eqref{eq:Fisher_matrix}, and \eqref{eq:gradient_omega}, $\textbf{C}_0$ and $\nabla \textbf{f}_0$ have already been computed for latent function inference with {\selfi}. 
The only missing quantity is the second matrix in the right-hand side of equation \eqref{eq:gradient_omega}, that is $\nabla_{\boldsymbol{\upomega}} \mathpzc{T}_0$, the gradient of $\mathpzc{T}$ evaluated at $\boldsymbol{\upomega}_0$. If unknown, its computation (e.g. via finite differences) does not require any more simulation of $\boldsymbol{\Phi}$. 
It is usually easy, as there are only $N$ directions in parameter space and $\mathpzc{T}$ is the numerically cheap part of the BHM. 
We note that, since we have to calculate $\textbf{F}_0$, we can easily get the Fisher-Rao distance between any simulated summaries $\widetilde{\boldsymbol{\upomega}}$ and the observed summaries $\widetilde{\boldsymbol{\upomega}}_\mathrm{O}$,
\begin{equation}
d_\mathrm{FR}(\widetilde{\boldsymbol{\upomega}}, \widetilde{\boldsymbol{\upomega}}_\mathrm{O}) \equiv \sqrt{\left(\widetilde{\boldsymbol{\upomega}}-\widetilde{\boldsymbol{\upomega}}_\mathrm{O}\right)^\intercal \textbf{F}_0 (\widetilde{\boldsymbol{\upomega}}-\widetilde{\boldsymbol{\upomega}}_\mathrm{O})},
\label{eq:FisherRao_distance}
\end{equation}
which can be used by any non-parametric SBI method.

We specify a prior $\p(\boldsymbol{\upomega})$ (typically peaking at or centred on $\boldsymbol{\upomega}_0$, for consistency with the assumptions made for data compression). 
Having defined $\mathpzc{C}$, we now have a full BHM that maps $\boldsymbol{\upomega}$ (of dimension $N$) to compressed summaries $\widetilde{\boldsymbol{\upomega}}$ (of size $N$), and that has been checked for model misspecification for the part linking $\boldsymbol{\uptheta}$ to $\boldsymbol{\Phi}$. 
We can then proceed with SBI via usual techniques. These can include likelihood-free rejection sampling, but also more sophisticated techniques such as \textsc{delfi} \citep[e.g.][]{Papamakarios2016,Alsing2018} or \textsc{bolfi} \citep[e.g.][]{GutmannCorander2016,Leclercq2018BOLFI,Thomas2020}.

\section{Lotka-Volterra BHM}
\label{sec:Lotka-Volterra BHM}

\subsection{Lotka-Volterra solver}

The Lotka-Volterra equations describe the dynamics of an ecological system in which two species interact, as a pair of first order non-linear differential equations:
\begin{eqnarray}
\deriv{x}{t} & = & \alpha x - \beta x y,\\
\deriv{y}{t} & = & \delta x y - \gamma y.
\end{eqnarray}
where $x(t)$ is the number of prey at time $t$ and $y(t)$ is the number of predators at time $t$. The model is characterised by $\boldsymbol{\upomega} = (\alpha, \beta, \gamma, \delta)$, a vector of four real parameters describing the interaction of the two species.

The initial conditions of the problem $\left\lbrace x(0),y(0) \right\rbrace = \left\lbrace x_0,y_0 \right\rbrace$ are assumed to be exactly known. Throughout the paper, timestepping and number of timesteps are fixed: $t_i = i \Delta t$ for $i \in \llbracket 0, S/2 \llbracket$. 

$\mathpzc{T}$ is an algorithm that numerically solves the ordinary differential equations. For simplicity, we choose an explicit Euler method: for all $i \in \llbracket 0, S/2 -1 \llbracket$,
\begin{eqnarray}
x(t_{i+1}) & = & x(t_i) \times \left[1 + \alpha - \beta y(t_i)\right] \times \Delta_t,\\
y(t_{i+1}) & = & y(t_i) \times \left[1 + \delta x(t_i) - \gamma \right] \times \Delta_t.
\end{eqnarray}

The latent function $\theta(t)$ is a concatenation of $x(t)$ and $y(t)$ evaluated at the timesteps of the problem. The corresponding vector is $\boldsymbol{\uptheta} \equiv \left\lbrace \left\lbrace x(t_i) \right\rbrace_{0 \leq i < S/2}, \left\lbrace y(t_i) \right\rbrace_{0 \leq i < S/2} \right\rbrace$ of size $S$.

\subsection{Lotka-Volterra observer}

\subsubsection{Full data model}
\label{sssec:Full data model}

To go from $\boldsymbol{\uptheta}$ to $\boldsymbol{\Phi}$, we assume a complex, probabilistic observational process of prey and predator populations, later referred to as ``model A'' and defined as follows.

\small{\textbf{\textsf{Signal.}}}\normalsize\ The (unobserved) signal $s_z$ is a delayed and non-linearly perturbed observation of the true population function for species $z \in \left\lbrace x,y \right\rbrace$, modulated by some seasonal efficiency $e_z(t)$. Formally, $s_x(0)=x_0$, $s_y(0)=y_0$, and for $i \in \llbracket 0, S/2-1 \llbracket$,
\begin{eqnarray}
s_x(t_{i+1}) & = & e_x(t_i) \left[ x(t_i) - p x(t_i) y(t_i) + q x(t_i)^2 \right]\!,\\
s_y(t_{i+1}) & = & e_y(t_i) \left[ y(t_i) + p x(t_i) y(t_i) - q y(t_i)^2 \right]\!.
\end{eqnarray}
These equations involve two parameters: $p$ accounts for hunts between $t_i$ and $t_{i+1}$ (making temporarily prey more likely to hide and predators more likely to be visible), and $q$ accounts for the gregariousness of prey and independence of predators. The free functions $e_x(t)$ and $e_y(t)$, valued in $[0,1]$, describe how prey and predators are likely to be detectable at any time, accounting for example for seasonal variation (hibernation, migration).

\small\textbf{\textsf{Noise.}}\normalsize\ The signal $s_z$ is subject to additive noise, giving a noisy signal $u_z(t) = s_z(t) + n^\mathrm{D}_z(t) + n^\mathrm{O}_z(t)$, where the noise has two components:
\begin{itemize}
\item demographic Gaussian noise with zero mean and variance proportional to the true underlying population, i.e. $n^\mathrm{D}_x(t) \curvearrowleft \mathpzc{G}\left[0,rx(t)\right]$ and $n^\mathrm{D}_y(t) \curvearrowleft \mathpzc{G}\left[0,ry(t)\right]$. The parameter $r$ gives the strength of demographic noise.
\item observational Gaussian noise that accounts for observer efficiency, coupling prey and predators such that
\begin{equation}
\begin{pmatrix}
n^\mathrm{O}_x(t) \\
n^\mathrm{O}_y(t)
\end{pmatrix}
\curvearrowleft \mathpzc{G} \left[
\begin{pmatrix}
0 \\
0
\end{pmatrix},
s \begin{pmatrix}
y(t) & t \sqrt{x(t)y(t)} \\
t \sqrt{x(t)y(t)} & x(t)
\end{pmatrix}
\right].
\label{eq:observational_noise}
\end{equation}
The parameter $s$ gives the overall amplitude of observational noise, and the parameter $t$ controls the strength of the non-diagonal component (it should be chosen such that the covariance matrix appearing in equation \eqref{eq:observational_noise} is positive semi-definite).
\end{itemize}

\small\textbf{\textsf{Censoring.}}\normalsize\ Finally, observed data are a censored and thresholded version of the noisy signal: for each timestep $t_i$, $\Phi_z(t_i) = m_z(t_i) \times \min\left[ u_z(t_i) , M_z \right]$ where $M_z$ is the maximum number of prey or predator that can be detected by the observer, and $m_z$ is a mask (taking either the value $0$ or $1$). Masked data points are discarded. The data vector is $\boldsymbol{\Phi} = \left\lbrace \left\lbrace \Phi_x(t_i) \right\rbrace, \left\lbrace \Phi_y(t_i) \right\rbrace \right\rbrace$. It contains $P \leq S$ elements depending on the number of masked timesteps for each species $z$ (formally, $P = \sum_{i=0}^{S/2-1} \left(\updelta_\mathrm{K}^{m_x(t_i),1} + \updelta_\mathrm{K}^{m_y(t_i),1} \right)$, where $\updelta_\mathrm{K}$ is a Kronecker delta symbol).

All of the free parameters ($p$, $q$, $r$, $s$, $t$, $M_x$, $M_y$) and free functions ($e_x(t)$, $e_y(t)$, $m_x(t)$, $m_y(t)$) appearing in the Lotka-Volterra observer data model described in this section are assumed known and fixed throughout the paper. 
Parameters used are $x_0=10$, $y_0=5$, $p=0.05$, $q=0.01$, $r=0.15$, $s=0.05$, $t=0.2$. 

\subsubsection{Simplified data model}
\label{sssec:Simplified data model}

In this section, we introduce ``model B'', a simplified (misspecified) data model linking $\boldsymbol{\uptheta}$ to $\boldsymbol{\Phi}$. Model B assumes that underlying functions are directly observed, i.e. $s_z(t) = z(t)$. It omits observational noise, such that $u_z(t) = s_z(t) + n^\mathrm{D}_z(t)$. In model B, parameters $p$, $q$, $s$ and $t$ are not involved, and the value of $r$ (strength of demographic noise) can be incorrect (we used $r=0.105$). Finally, model B fails to account for the thresholds: $\Phi_z(t) = m_z(t) u_z(t)$.

\section{Results}
\label{sec:Results}

In this section, we apply the two-step inference method described in section \ref{sec:Method} to the Lotka-Volterra BHM introduced in section \ref{sec:Lotka-Volterra BHM}. We generate mock data $\boldsymbol{\Phi}_\mathrm{O}$ from model A, using ground truth parameters $\boldsymbol{\upomega}_\mathrm{gt} = (\alpha_\mathrm{gt}, \beta_\mathrm{gt}, \gamma_\mathrm{gt}, \delta_\mathrm{gt}) = (0.55,0.2,0.2,0.05)$. 
We assume that ground truth parameters are known \textit{a priori} with a precision of about $3\%$. Consistently, we choose a Gaussian prior $\p(\boldsymbol{\upomega})$ with mean $\boldsymbol{\upomega}_0 = (0.5768, 0.1963, 0.1968, 0.0484)$ and diagonal covariance matrix $\mathrm{diag}(0.0173^2, 0.0059^2, 0.0059^2, 0.0015^2)$.

\subsection{Inference of population functions with SELFI}

\begin{figure*}
\begin{center}
\includegraphics[width=\textwidth]{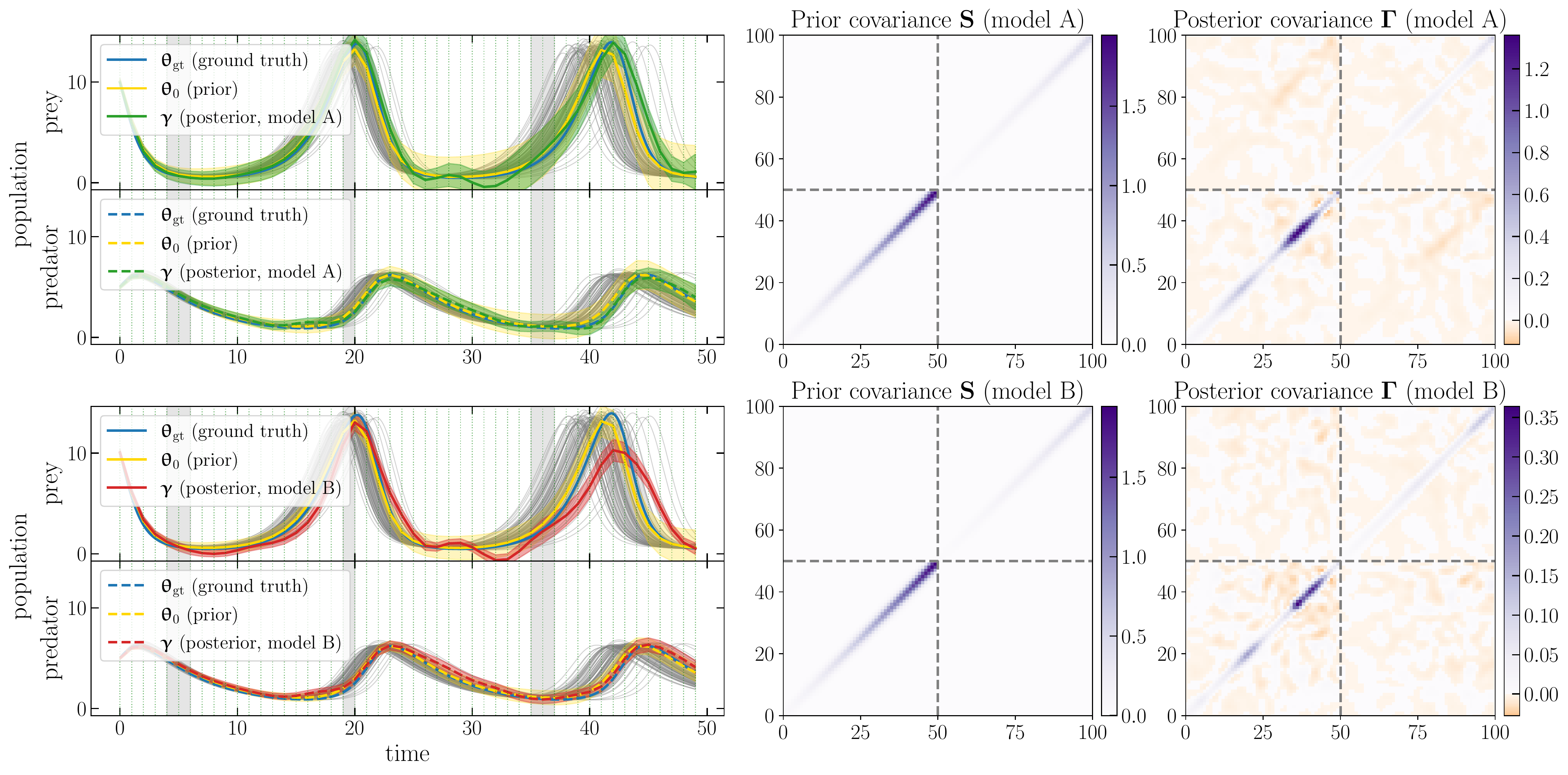}
\caption{\label{fig:selfi}{\selfi} inference of the population function $\boldsymbol{\uptheta}$ given the observed data $\boldsymbol{\Phi}_\mathrm{O}$, used as a check for model misspecification. \textit{Left panels}. The prior mean and expansion point $\boldsymbol{\uptheta}_0$ and the effective posterior mean $\boldsymbol{\upgamma}$ are represented as yellow and green/red lines, respectively, with their $2\sigma$ credible intervals. For comparison, simulations $\mathpzc{T}(\boldsymbol{\upomega})$ with $\boldsymbol{\upomega} \curvearrowleft \p(\boldsymbol{\upomega})$, and the ground truth $\boldsymbol{\uptheta}_\mathrm{gt}$ are shown in grey and blue, respectively. \textit{Middle and right panels}. The prior covariance matrix $\textbf{S}$ and the posterior covariance matrix $\boldsymbol{\Gamma}$, respectively. The first row corresponds to model A (see section \ref{sssec:Full data model}) and the second row to model B (see section \ref{sssec:Simplified data model}).}
\end{center}
\end{figure*}

We first seek to reconstruct the latent population functions $x(t)$ and $y(t)$, conditional on the data $\boldsymbol{\Phi}_\mathrm{O}$, using {\selfi}. 
We choose as expansion point the population functions simulated from the mean of the prior on $\boldsymbol{\upomega}$, i.e. $\boldsymbol{\uptheta}_0 = \mathpzc{T}(\boldsymbol{\upomega}_0)$.
We use $N_0 = 150$ and $N_s = 100$; the computational workload is therefore a fixed number of $10,150$ simulations for each model. It is known \textit{a priori} and perfectly parallel. 

We adopt a Gaussian prior $\p(\boldsymbol{\uptheta})$ and combine it with the effective likelihood to obtain the {\selfi} effective posterior $\p(\boldsymbol{\uptheta}|\boldsymbol{\Phi}_\mathrm{O})$. 
Figure \ref{fig:selfi} (left panels) shows the inferred population functions $\boldsymbol{\upgamma}$ in comparison with the prior mean and expansion point $\boldsymbol{\uptheta}_0$ and the ground truth $\boldsymbol{\uptheta}_\mathrm{gt}$. $2\sigma$ credible regions are shown for the prior and the posterior (i.e. $2\sqrt{\mathrm{diag}(\textbf{S})}$ and $2\sqrt{\mathrm{diag}(\boldsymbol{\Gamma})}$, respectively). 
The full posterior covariance matrix $\boldsymbol{\Gamma}$ for each model is shown in the rightmost column of figure \ref{fig:selfi}.

\subsection{Check for model misspecification}
\label{ssec:Check for model misspecification}

The inferred population functions allow us to check for model misspecification. 
From figure \ref{fig:selfi}, it is clear that model B fails to produce a plausible reconstruction of population functions: model B breaks the \mbox{(pseudo-)periodicity} of the predator population function $y(t)$, which is a property required by the model. 
In the bottom left-hand panels, the red lines differ in shape from fiducial functions $\mathpzc{T}(\boldsymbol{\upomega})$ (grey lines), and the credible intervals exclude the expansion point. 
On the contrary, with model A, the reconstructed population functions are consistent with the expansion point. The inference is unbiased, since the ground truth typically lies within the $2\sigma$ credible region of the reconstruction. 

As a quantitative check, we compute the Mahalanobis distance between $\boldsymbol{\upgamma}$ and $\p(\boldsymbol{\uptheta})$ (equation \eqref{eq:Mahalanobis_distance_mscheck}) for each model. We find that $d_\mathrm{M}(\boldsymbol{\upgamma},\boldsymbol{\uptheta}_0|\textbf{S})$ is much smaller for model A than for model B ($5.35$ versus $12.54$). The numbers can be compared to the empirical mean among our set of fiducial populations functions, $\left\langle d_\mathrm{M}(\mathpzc{T}(\boldsymbol{\upomega}_n), \boldsymbol{\uptheta}_0 | \textbf{S}) \right\rangle = 9.43$.

At this stage, we therefore consider that model B is excluded, and we proceed further with model A.

\subsection{Score compression}

\begin{figure*}
\begin{center}
\vspace*{-5pt}
\includegraphics[width=\textwidth]{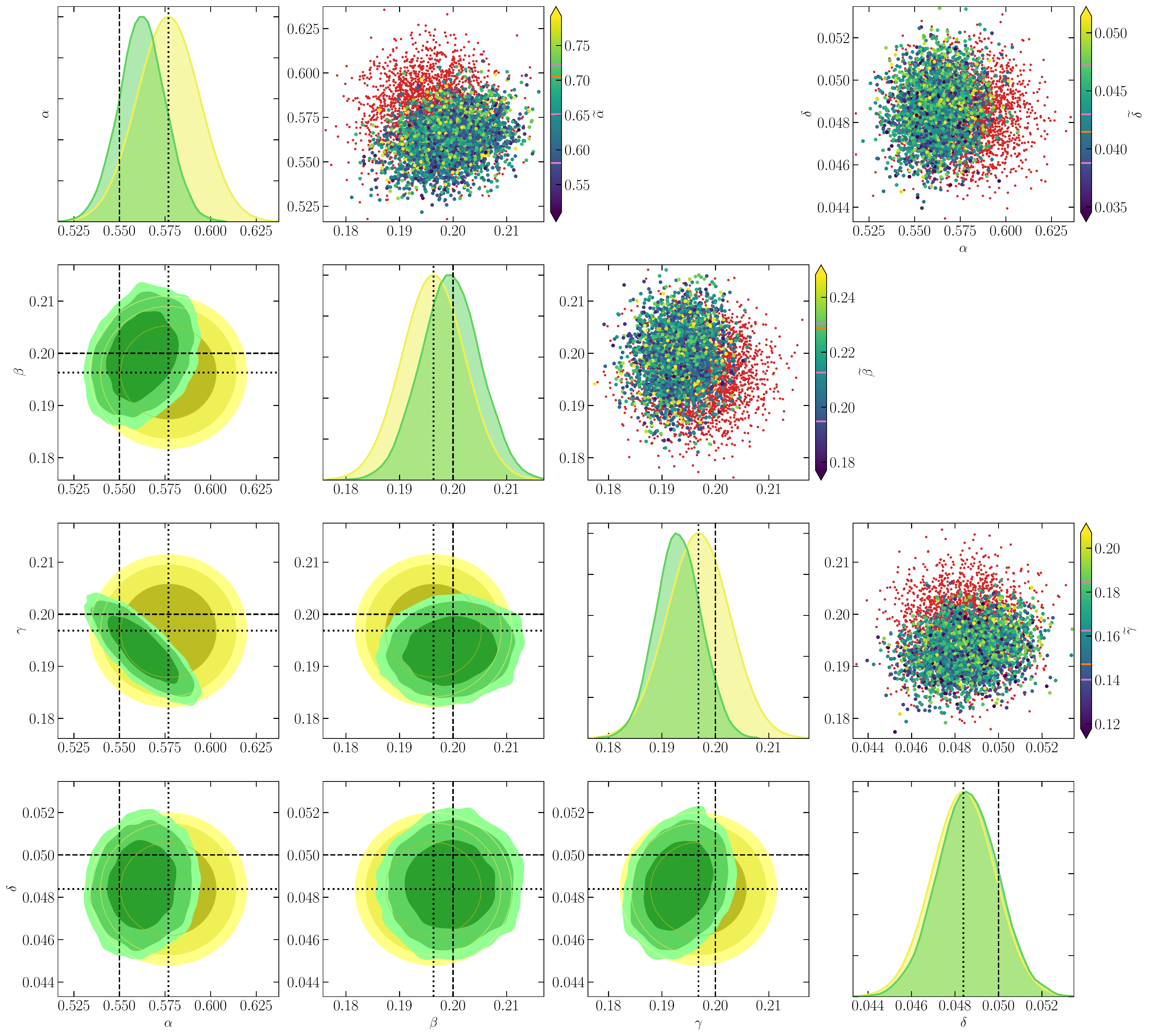}
\vspace*{-20pt}
\caption{\label{fig:abc}Simulation-based inference of the Lotka-Volterra parameters $\boldsymbol{\upomega}=(\alpha,\beta,\gamma,\delta)$ given the compressed observed data $\boldsymbol{\widetilde{\upomega}}_\mathrm{O}$. Plots in the lower corner show two-dimensional marginals of the prior $\p(\boldsymbol{\upomega})$ (yellow contours) and of the SBI posterior $\p(\boldsymbol{\upomega}|\boldsymbol{\widetilde{\upomega}}_\mathrm{O})$ (green contours), using a threshold $\varepsilon = 2$ on the Fisher-Rao distance between simulated $\boldsymbol{\widetilde{\upomega}}$ and observed $\boldsymbol{\widetilde{\upomega}}_\mathrm{O}$, $d_\mathrm{FR}(\boldsymbol{\widetilde{\upomega}}, \boldsymbol{\widetilde{\upomega}}_\mathrm{O})$. Contours show $1$, $2$, and $3\sigma$ credible regions. Plots on the diagonal show one-dimensional marginal distributions of the parameters, using the same colour scheme. Dotted and dashed lines denote the position of the fiducial point for score compression $\boldsymbol{\upomega}_0$ and of the ground truth parameters $\boldsymbol{\upomega}_\mathrm{gt}$, respectively. 
The scatter plots in the upper corner illustrate score compression for pairs of parameters. There, red dots represent some simulated samples. Larger dots show some accepted samples (i.e. for which $d_\mathrm{FR}(\boldsymbol{\widetilde{\upomega}}, \boldsymbol{\widetilde{\upomega}}_\mathrm{O}) < \varepsilon$), with a colour map corresponding to the value of one component of $\boldsymbol{\widetilde{\upomega}}$. In the colour bars, pink lines denote the mean and $1\sigma$ scatter among accepted samples of the component of $\boldsymbol{\widetilde{\upomega}}$, and the orange line denotes its value in $\boldsymbol{\widetilde{\upomega}}_\mathrm{O}$.
\vspace*{-15pt}
}
\end{center}
\end{figure*}

As $\mathpzc{T}$ is numerically cheap, we get $\nabla_{\boldsymbol{\upomega}} \mathpzc{T}_0$ via sixth-order central finite differences around $\boldsymbol{\upomega}_0$, then obtain $\nabla_{\boldsymbol{\upomega}} \textbf{f}_0$ using equation \eqref{eq:gradient_omega}. This does not require any further evaluation of the data model $\p(\boldsymbol{\Phi}|\boldsymbol{\uptheta})$, since $\nabla \textbf{f}_0$ has already been computed.

Using equations \eqref{eq:compression_mle} and \eqref{eq:Fisher_matrix}, we compress $\boldsymbol{\Phi}_\mathrm{O}$ and obtain $\boldsymbol{\widetilde{\upomega}}_\mathrm{O} = (0.7050, 0.2287, 0.1471, 0.0415)$.

\subsection{Inference of parameters using likelihood-free rejection sampling}

As a last step, we infer top-level parameters $\boldsymbol{\upomega}$ given compressed summaries $\boldsymbol{\widetilde{\upomega}}_\mathrm{O}$. As the problem studied in this paper is sufficiently simple, we rely on the simplest solution for SBI, namely likelihood-free rejection sampling \citep[sometimes also known as Approximate Bayesian Computation, e.g.][]{Beaumont2019}. To do so, we use the Fisher-Rao distance between simulated $\boldsymbol{\widetilde{\upomega}}$ and observed $\boldsymbol{\widetilde{\upomega}}_\mathrm{O}$, which comes naturally from score compression (see equation \eqref{eq:FisherRao_distance}), and we set a threshold $\varepsilon = 2$. We draw samples from the prior $\p(\boldsymbol{\upomega})$, simulate $\boldsymbol{\widetilde{\upomega}}$, then accept $\boldsymbol{\upomega}$ as a sample of $\p(\boldsymbol{\upomega}|\boldsymbol{\widetilde{\upomega}}_\mathrm{O})$ if $d_\mathrm{FR}(\boldsymbol{\widetilde{\upomega}}, \boldsymbol{\widetilde{\upomega}}_\mathrm{O}) < \varepsilon$, and reject it otherwise. 

In figure \ref{fig:abc}, we find that the inference of top-level parameters is unbiased, with the ground truth $\boldsymbol{\upomega}_\mathrm{gt}$ (dashed lines) lying within the $2\sigma$ credible region of the posterior. We observe that the data correctly drive some features that are not built into the prior, for instance, the degeneracy between $\alpha$ and $\gamma$, respectively the reproduction rate of prey and the mortality rate of predators. 

\section{Conclusion}
\label{sec:Conclusion}

One of the biggest challenges in statistical data analysis is checking data models for misspecification, so as to obtain meaningful parameter inferences. 
In this work, we described a novel two-step simulation-based Bayesian approach, combining {\selfi} and SBI, which can be used to tackle this issue for a large class of models. 
BHMs to which the approach can be applied involve a latent function depending on parameters and observed through a complex probabilistic process. 
They are ubiquitous, e.g. in astrophysics and ecology.

In this paper, we introduced a prey-predator model, consisting of a numerical solver of the Lotka-Volterra system of equations and of a complex observational process of population functions. As a proof of concept, we applied our technique to this model and to a simplified (misspecified) version of it. 
We demonstrated successful identification of the misspecified model and unbiased inference of the parameters of the correct model.

In conclusion, the method developed constitutes a computationally efficient and easily applicable framework to perform SBI of BHMs while checking for model misspecification. 
It allows one to infer the latent function as an intermediate product, then to perform score compression at no additional simulation cost. 
This study opens up a new avenue to increase the robustness and reliability of Bayesian data analysis using fully non-linear, simulator-based models.

The code and data underlying this paper, as well as additional plots, will be made publicly available as part of the py{\selfi} code at \href{https://pyselfi.florent-leclercq.eu}{https://pyselfi.florent-leclercq.eu}.

\section*{References}
\bibliography{MaxEnt2022_paper}

\begin{thebibliography}{9}%
\makeatletter
\providecommand \@ifxundefined [1]{%
 \@ifx{#1\undefined}
}%
\providecommand \@ifnum [1]{%
 \ifnum #1\expandafter \@firstoftwo
 \else \expandafter \@secondoftwo
 \fi
}%
\providecommand \@ifx [1]{%
 \ifx #1\expandafter \@firstoftwo
 \else \expandafter \@secondoftwo
 \fi
}%
\providecommand \natexlab [1]{#1}%
\providecommand \enquote  [1]{``#1''}%
\providecommand \bibnamefont  [1]{#1}%
\providecommand \bibfnamefont [1]{#1}%
\providecommand \citenamefont [1]{#1}%
\providecommand \href@noop [0]{\@secondoftwo}%
\providecommand \href [0]{\begingroup \@sanitize@url \@href}%
\providecommand \@href[1]{\@@startlink{#1}\@@href}%
\providecommand \@@href[1]{\endgroup#1\@@endlink}%
\providecommand \@sanitize@url [0]{\catcode `\\12\catcode `\$12\catcode
  `\&12\catcode `\#12\catcode `\^12\catcode `\_12\catcode `\%12\relax}%
\providecommand \@@startlink[1]{}%
\providecommand \@@endlink[0]{}%
\newcommand{\PineGreen}[1]{\textcolor{PineGreen}{#1}}%
\providecommand \url  [0]{\begingroup\@sanitize@url \@url }%
\providecommand \@url [1]{\endgroup\@href {#1}{\urlprefix }}%
\providecommand \urlprefix  [0]{URL }%
\providecommand \Eprint [0]{\href }%
\providecommand \doibase [0]{http://dx.doi.org/}%
\providecommand \selectlanguage [0]{\@gobble}%
\providecommand \bibinfo  [0]{\@secondoftwo}%
\providecommand \bibfield  [0]{\@secondoftwo}%
\providecommand \translation [1]{[#1]}%
\providecommand \BibitemOpen [0]{}%
\providecommand \bibitemStop [0]{}%
\providecommand \bibitemNoStop [0]{.\EOS\space}%
\providecommand \EOS [0]{\spacefactor3000\relax}%
\providecommand \BibitemShut  [1]{\csname bibitem#1\endcsname}%
\let\auto@bib@innerbib\@empty
\bibitem [{Alsing \& Wandelt(2018)\citenamefont {Alsing\ \&\
  Wandelt}}]{AlsingWandelt2018}%
{(\PineGreen{Alsing \& Wandelt}, \PineGreen{2018})}  \BibitemOpen
  \bibfield  {author} {\bibinfo {author} {\bibfnamefont {J.}~\bibnamefont
  {Alsing}}, \bibinfo {author} {\bibfnamefont {B.}~\bibnamefont {Wandelt}},\
  }\emph {Generalized massive optimal data compression},\ \href {\doibase
  10.1093/mnrasl/sly029} {\bibfield  {journal} {\bibinfo  {journal} {Monthly
  Notices of the Royal Astronomical Society Letters}\ }\textbf {\bibinfo
  {volume} {476}},\ \bibinfo {pages} {L60} (\bibinfo {year} {2018})},\ \Eprint
  {https://arxiv.org/abs/1712.00012} {arXiv:1712.00012 [astro-ph.CO]}
  \BibitemShut {NoStop}%
\bibitem [{Alsing, Wandelt \& Feeney(2018)\citenamefont {Alsing, Wandelt,\ \&\
  Feeney}}]{Alsing2018}%
{(\PineGreen{Alsing, Wandelt \& Feeney}, \PineGreen{2018})}  \BibitemOpen
  \bibfield  {author} {\bibinfo {author} {\bibfnamefont {J.}~\bibnamefont
  {Alsing}}, \bibinfo {author} {\bibfnamefont {B.}~\bibnamefont {Wandelt}},
  \bibinfo {author} {\bibfnamefont {S.}~\bibnamefont {Feeney}},\ }\emph
  {Massive optimal data compression and density estimation for scalable,
  likelihood-free inference in cosmology},\ \href {\doibase
  10.1093/mnras/sty819} {\bibfield  {journal} {\bibinfo  {journal} {Monthly
  Notices of the Royal Astronomical Society}\ }\textbf {\bibinfo {volume}
  {477}},\ \bibinfo {pages} {2874} (\bibinfo {year} {2018})},\ \Eprint
  {https://arxiv.org/abs/1801.01497} {arXiv:1801.01497} \BibitemShut {NoStop}%
\bibitem [{Beaumont(2019)\citenamefont {Beaumont}}]{Beaumont2019}%
{(\PineGreen{Beaumont}, \PineGreen{2019})}  \BibitemOpen
  \bibfield  {author} {\bibinfo {author} {\bibfnamefont {M.~A.}\ \bibnamefont
  {Beaumont}},\ }\emph {Approximate Bayesian Computation},\ \href {\doibase
  10.1146/annurev-statistics-030718-105212} {\bibfield  {journal} {\bibinfo
  {journal} {Annual Review of Statistics and Its Application}\ }\textbf
  {\bibinfo {volume} {6}},\ \bibinfo {pages} {379} (\bibinfo {year} {2019})},\
  \Eprint
  {https://arxiv.org/abs/https://doi.org/10.1146/annurev-statistics-030718-105212}
  {https://doi.org/10.1146/annurev-statistics-030718-105212} \BibitemShut
  {NoStop}%
\bibitem [{{Gutmann} \& {Corander}(2016)\citenamefont {{Gutmann}\ \&\
  {Corander}}}]{GutmannCorander2016}%
{(\PineGreen{{Gutmann} \& {Corander}}, \PineGreen{2016})}  \BibitemOpen
  \bibfield  {author} {\bibinfo {author} {\bibfnamefont {M.~U.}\ \bibnamefont
  {{Gutmann}}}, \bibinfo {author} {\bibfnamefont {J.}~\bibnamefont
  {{Corander}}},\ }\emph {{Bayesian Optimization for Likelihood-Free Inference
  of Simulator-Based Statistical Models}},\ \href
  {http://jmlr.org/papers/v17/15-017.html} {\bibfield  {journal} {\bibinfo
  {journal} {Journal of Machine Learning Research}\ }\textbf {\bibinfo {volume}
  {17}},\ \bibinfo {pages} {1} (\bibinfo {year} {2016})},\ \Eprint
  {https://arxiv.org/abs/1501.03291} {arXiv:1501.03291 [stat.ML]} \BibitemShut
  {NoStop}%
\bibitem [{{Leclercq}(2018)\citenamefont {{Leclercq}}}]{Leclercq2018BOLFI}%
{(\PineGreen{{Leclercq}}, \PineGreen{2018})}  \BibitemOpen
  \bibfield  {author} {\bibinfo {author} {\bibfnamefont {F.}~\bibnamefont
  {{Leclercq}}},\ }\emph {{Bayesian optimization for likelihood-free
  cosmological inference}},\ \href {\doibase 10.1103/PhysRevD.98.063511}
  {\bibfield  {journal} {\bibinfo  {journal} {Physical Review D}\ }\textbf
  {\bibinfo {volume} {98}},\ \bibinfo {eid} {063511} (\bibinfo {year}
  {2018})},\ \Eprint {https://arxiv.org/abs/1805.07152} {arXiv:1805.07152
  [astro-ph.CO]} \BibitemShut {NoStop}%
\bibitem [{{Leclercq} {\textit{et~al}}\mbox{.}(2019)\citenamefont {{Leclercq},
  {Enzi}, {Jasche},\ \&\ {Heavens}}}]{Leclercq2019SELFI}%
{(\PineGreen{{Leclercq} {\textit{et~al}}\mbox{.}}, \PineGreen{2019})}
  \BibitemOpen
  \bibfield  {author} {\bibinfo {author} {\bibfnamefont {F.}~\bibnamefont
  {{Leclercq}}}, \bibinfo {author} {\bibfnamefont {W.}~\bibnamefont {{Enzi}}},
  \bibinfo {author} {\bibfnamefont {J.}~\bibnamefont {{Jasche}}}, \bibinfo
  {author} {\bibfnamefont {A.}~\bibnamefont {{Heavens}}},\ }\emph {{Primordial
  power spectrum and cosmology from black-box galaxy surveys}},\ \href
  {\doibase 10.1093/mnras/stz2718} {\bibfield  {journal} {\bibinfo  {journal}
  {Monthly Notices of the Royal Astronomical Society}\ }\textbf {\bibinfo
  {volume} {490}},\ \bibinfo {pages} {4237} (\bibinfo {year} {2019})},\ \Eprint
  {https://arxiv.org/abs/1902.10149} {arXiv:1902.10149 [astro-ph.CO]}
  \BibitemShut {NoStop}%
\bibitem [{{Papamakarios} \& {Murray}(2016)\citenamefont {{Papamakarios}\ \&\
  {Murray}}}]{Papamakarios2016}%
{(\PineGreen{{Papamakarios} \& {Murray}}, \PineGreen{2016})}  \BibitemOpen
  \bibfield  {author} {\bibinfo {author} {\bibfnamefont {G.}~\bibnamefont
  {{Papamakarios}}}, \bibinfo {author} {\bibfnamefont {I.}~\bibnamefont
  {{Murray}}},\ }\emph {{Fast $\epsilon$-free Inference of Simulation Models
  with Bayesian Conditional Density Estimation}},\ \href
  {http://papers.nips.cc/paper/6084-fast-free-inference-of-simulation-models-with-bayesian-conditional-density-estimation.pdf}
  {\bibfield  {journal} {\bibinfo  {journal} {Advances in Neural Information
  Processing Systems}\ }\textbf {\bibinfo {volume} {29}} (\bibinfo {year}
  {2016})},\ \Eprint {https://arxiv.org/abs/1605.06376} {arXiv:1605.06376
  [stat.ML]} \BibitemShut {NoStop}%
\bibitem [{Rousset(2007)\citenamefont {Rousset}}]{Rousset2007}%
{(\PineGreen{Rousset}, \PineGreen{2007})}  \BibitemOpen
  \bibfield  {author} {\bibinfo {author} {\bibfnamefont {F.}~\bibnamefont
  {Rousset}},\ }\emph {Inferences from Spatial Population Genetics},\ in\ \href
  {\doibase https://doi.org/10.1002/9780470061619.ch28} {\emph {\bibinfo
  {booktitle} {Handbook of Statistical Genetics}}}\ (\bibinfo  {publisher}
  {John Wiley \& Sons, Ltd},\ \bibinfo {year} {2007})\ Chap.~\bibinfo {chapter}
  {28}, pp.\ \bibinfo {pages} {945--979}\BibitemShut {NoStop}%
\bibitem [{Thomas {\textit{et~al}}\mbox{.}(2020)\citenamefont {Thomas, Pesonen,
  S{\'a}-Le{\~a}o, de~Lencastre, Kaski,\ \&\ Corander}}]{Thomas2020}%
{(\PineGreen{Thomas {\textit{et~al}}\mbox{.}}, \PineGreen{2020})}  \BibitemOpen
  \bibfield  {author} {\bibinfo {author} {\bibfnamefont {O.}~\bibnamefont
  {Thomas}}, \bibinfo {author} {\bibfnamefont {H.}~\bibnamefont {Pesonen}},
  \bibinfo {author} {\bibfnamefont {R.}~\bibnamefont {S{\'a}-Le{\~a}o}},
  \bibinfo {author} {\bibfnamefont {H.}~\bibnamefont {de~Lencastre}}, \bibinfo
  {author} {\bibfnamefont {S.}~\bibnamefont {Kaski}}, \bibinfo {author}
  {\bibfnamefont {J.}~\bibnamefont {Corander}},\ }\emph {Split-BOLFI for
  misspecification-robust likelihood free inference in high dimensions},\ \href
  {https://ui.adsabs.harvard.edu/abs/2020arXiv200209377T} {\bibfield  {journal}
  {\bibinfo  {journal} {arXiv e-prints}\ } (\bibinfo {year} {2020})},\ \Eprint
  {https://arxiv.org/abs/2002.09377} {arXiv:2002.09377 [stat.CO]} \BibitemShut
  {NoStop}%
\end{thebibliography}%

\appendix

\end{document}